\documentclass[10pt]{iopart}

\usepackage[utf8]{inputenc}
\expandafter\let\csname equation*\endcsname\relax
\expandafter\let\csname endequation*\endcsname\relax
\usepackage{amsmath}

\usepackage{iopams}
\usepackage{xcolor}
\usepackage{ifthen}
\usepackage{siunitx}
\DeclareSIUnit\bar{bar}

\usepackage{braket}

\usepackage{tikz}
\usetikzlibrary{arrows}
\usetikzlibrary{calc}
\usetikzlibrary{decorations.pathreplacing,calligraphy}

\newcommand{\NiceCircleNumber} [1] {\raisebox{.5pt}{\textcircled{\raisebox{-.9pt} {#1}}}
}

\newcommand{\FigRef}[1]{figure~\ref{#1}}
\newcommand{\FIGRef}[1]{Figure~\ref{#1}}
\newcommand{\SubFigRef}[2]{figure~\ref{#1}(#2)}
\newcommand{\EqnRef}[1]{equation~\ref{#1}}
\newcommand{\TblRef}[1]{table~\ref{#1}}

\usepackage[english]{babel}
\usepackage{graphics}
\usepackage{todonotes}
\usepackage{graphicx}
\usepackage{comment}
\usepackage{xurl}
\usepackage{cite}

\newboolean{Anonymous}\setboolean{Anonymous}{false}
\provideboolean{Anonymous}
\newcommand{\Anonymize}[1]{%
\ifthenelse{\boolean{Anonymous}}{\emph{hidden text}}{#1}}

\begin{document}

\title[Doppler-free high resolution continuous wave spectroscopy on the $\mathrm{A}\,^2\Sigma^+ \leftarrow \mathrm{X}\,^2\Pi_{3/2}$ transition in nitric oxide]{Doppler-free high resolution continuous wave optical UV-spectroscopy on the $\mathrm{A}\,^2\Sigma^+ \leftarrow \mathrm{X}\,^2\Pi_{3/2}$ transition in nitric oxide}

\author{\Anonymize{Patrick Kaspar$^1$, Fabian Munkes$^1$, Philipp Neufeld$^1$, Lea Ebel$^1$, Yannick Schellander$^2$, Robert Löw$^1$, Tilman Pfau$^1$, and Harald Kübler$^1$}}

\address{\Anonymize{$^1$ University of Stuttgart, 5. Institute of Physics and Center for Integrated Quantum Science and Technology IQST, Pfaffenwaldring 57, 70569 Stuttgart, Germany}}
\address{\Anonymize{$^2$ University of Stuttgart, Institute for Large Area Microelectronics, Allmandring 3b, 70569 Stuttgart, Germany}}
\ead{\Anonymize{h.kuebler@physik.uni-stuttgart.de}}
\vspace{10pt}
\begin{indented}
\item[]June 2022
\end{indented}

\begin{abstract}
 We report on Doppler-free continuous-wave optical UV-spectroscopy resolving the hyperfine structure of the $\mathrm{A}\,^2\Sigma^+ \leftarrow \mathrm{X}\,^2\Pi_{3/2}$ transition in nitric oxide for total angular momenta $J_\mathrm{X}=1.5-19.5$ on the $\mathrm{oP_{12ee}}$ branch. The resulting line splittings are compared to calculated splittings and fitted determining new values for the molecular constants $b$, $c$, $eQq_0$ and $b_F$ for the $\mathrm{A}\,^2\Sigma^+$ state. The constants are in good agreement with values previously determined by quantum beat spectroscopy.

\end{abstract}
\ioptwocol
\section{Introduction}\label{Sec1}
Nitric oxide (NO) has been subject to a large number of spectroscopic studies. It is of great interest to radio astronomy and atmospheric science \cite{Altshuller1986,Crutzen1979}. Especially the $\mathrm{\gamma}$-band comprising the transitions between the two electronic states $\mathrm{A}\,^2\Sigma^+$ and $\mathrm{X}\,^2\Pi$ has been thoroughly investigated experimentally \cite{Hall1966,Engleman1969,Tajime1978,Ishii1994, Zobnin1995,Wang1996,Danielak1997} and theoretically \cite{Langhoff1988,Cheng2017,Polak2003}.
NO shows a hyperfine structure which is, for the most abundant isotope ($>99\,\%$) $^{14}\mathrm{N}^{16}\mathrm{O}$, due to the nuclear spin $I=1$ of nitrogen.   
The doublet ground state $\mathrm{X}\,^2\Pi$ shows large spin-orbit splitting and $\Lambda$-type doubling. The latter was subject to the studies of Neumann \cite{Neumann1970} and Paul \cite{Paul1997}.

For the energetically lowered spin-orbit component $\Pi_{1/2}$ the $\Lambda$-type doubling is larger and increases linearly with $J_\mathrm{X}$ while the splitting in the $\Pi_{3/2}$ component is smaller but increases proportional to $J_\mathrm{X}^2$.

The hyperfine structure of the ground state has been investigated with different spectroscopic techniques. Early investigations relied on microwave spectroscopy \cite{Beringer1954,Gallagher1956,Favero1959,Brown1966}. Meerts and Dymanus \cite{Meerts1972} resolved several hyperfine transitions for both ground state manifolds for different total angular momenta $J_\mathrm{X}$ and provided matrix elements to treat the hyperfine structure with perturbation theory up to the third order. Further theoretical work was provided by Mizushima \cite{Mizushima1954} and Kristiansen \cite{Kristiansen1977}. High precision infrared spectroscopy allowed the determination of hyperfine constants for the two spin-orbit components at the \SI{10}{\hertz}-level accuracy \cite{Saupe1996,Varberg1999}.

In the excited state $A\,^2\Sigma^+$ spin-rotational splitting occurs \cite{Timmermann1981,Wallenstein1978}. The hyperfine structure of the excited state was theoretically treated by Green \cite{Green1972,Green1973} and Walch \cite{Walch1975}. First experimental results were published by Bergemann and Zare employing radio-frequency double resonance \cite{Bergeman1974}. The $v\!=\!1$ vibrational level was resolved via two-photon spectroscopy yielding fine and hyperfine parameters \cite{Murphy1993,Miller1989}. In addition the electric dipole moment of the $\mathrm{A}\,^2\Sigma^+$ state was investigated experimentally \cite{Gray1993} and theoretically \cite{Glendening1995}. Reid observed hyperfine quantum beats with time-resolved photoelectron spectroscopy \cite{Reid1994} while further measurements relied on two-color resonant four-wave mixing \cite{McCormack2001} and on quantum beat spectroscopy \cite{McCormack1994,Brouard2012}.

Nitric oxide is involved in a multitude of chemical and biological processes in the human body \cite{Furchgott1980,Ignarro1987,Yun1996,Moncada2006}. It acts as an indicator for inflammatory diseases like asthma or cancer \cite{Alving1993,Lundberg1996,Mazzatenta2013}. Therefore, detecting small amounts of nitric oxide in a large background of other gases is of particular interest in medical research. A novel approach to detect nitric oxide relies on optogalvanic detection. An optical excitation to Rydberg states is followed by collisional ionization and electrical detection of the thereby generated charges \cite{Kaspar2021}. The working principle of this type of sensing scheme has already been demonstrated in an idealized system and a proof of concept study \cite{Schmidt2018,Schmidt2020}. Employing narrow-band lasers instead of pulsed systems for the excitation increases the selectivity of such a sensor.

Within the scope of the development of a laboratory prototype sensor relying on narrow-band lasers, Doppler-free saturation spectroscopy was performed to gain precise knowledge of the substructure of the involved states. The resolved individual Lamb-Dips, which were also recently observed for the HD molecule \cite{Tao2018,Diouf2019}, are essential to quantify and assess the capabilities of the sensor prototype.

Doppler-free continuous wave spectroscopy was performed on the $\mathrm{oP}_{12\mathrm{ee}}$ branch of the $\mathrm{A}\,^2\Sigma^+, (v\!=\!0) \leftarrow \mathrm{X}\,^2\Pi_{3/2}, (v\!=\!0)$ transition for total angular momenta $J_{\mathrm{X}}=1.5$ to $19.5$. The hyperfine structure has been partially resolved and corresponding hyperfine constants of the $\mathrm{A}\,^2\Sigma^+$ state are fitted and compared to previously determined values. For the fit only data for $J_{\mathrm{X}}>5.5$ is included.
\section{Experimental Methods}\label{Sec2}
\begin{figure}[h]
    \centering
    \begin{tikzpicture}
    \node[anchor=south west] (image) at (0,0){\includegraphics[width=0.45 \textwidth]{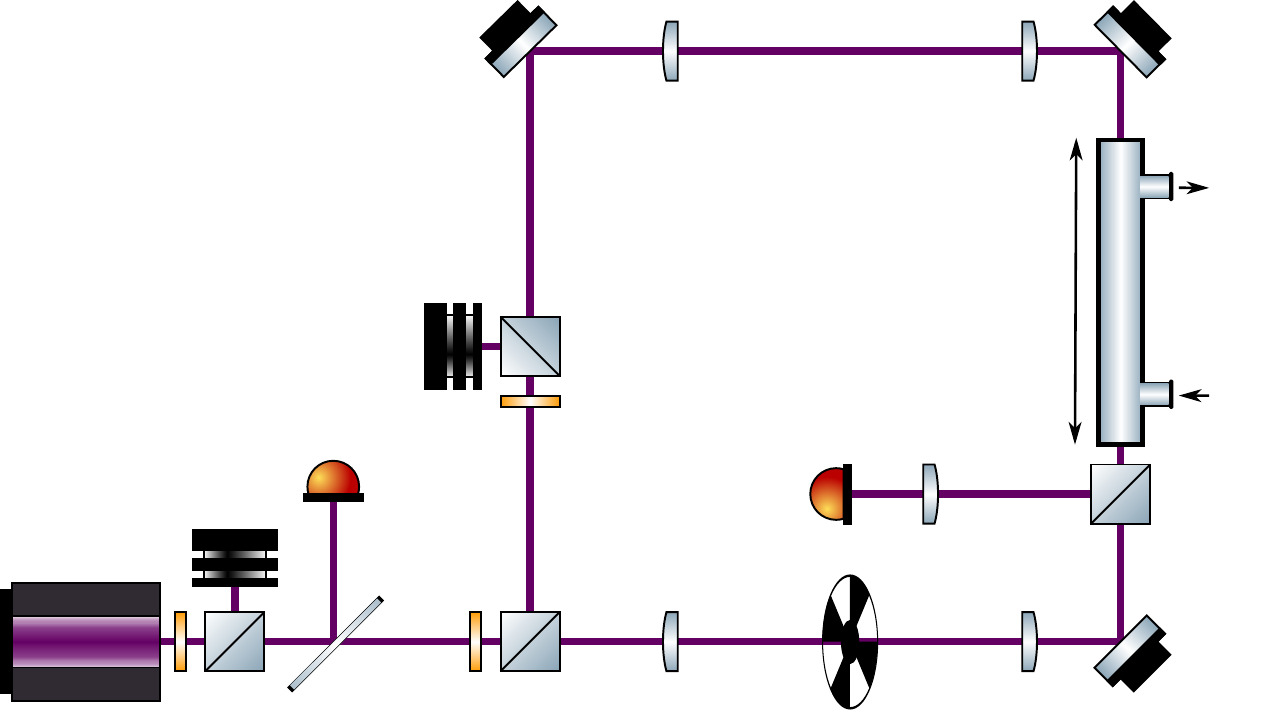}};
    \begin{scope}[
    x={($0.1*(image.south east)$)},
    y={($0.1*(image.north west)$)}]
    \node[text=black, anchor=center] at (6.6,-0.25) {Chopper Wheel};
    \node[text=black, anchor=west] at (-0.6,2.5) {UV-Laser}; 
    \node[text=black, anchor=center] at (2.6,-0.25) {Pick-Off};
    \node[text=black, anchor=center] at (7.5,6.3) {Cell};
    \node[text=black, anchor=center] at (7.5,5.5) {50 cm};
    \node[text=black] at (2.7,4) {PD1};
    \node[text=black] at (9.3,5.2) {In};
    \node[text=black] at (9.3,7.9) {Out};
    \node[text=black, anchor=center] at (5.7,3.2) {PD2};
    \end{scope}
    \end{tikzpicture}
    \caption{\label{setup} Schematic of the spectroscopy setup. The output of the UV-Laser is split into a pump and a probe beam which are both expanded before entering the cell in counter-propagating configuration. The pump beam is modulated with a chopper wheel to employ a lock-in amplifier. 
    }
\end{figure}
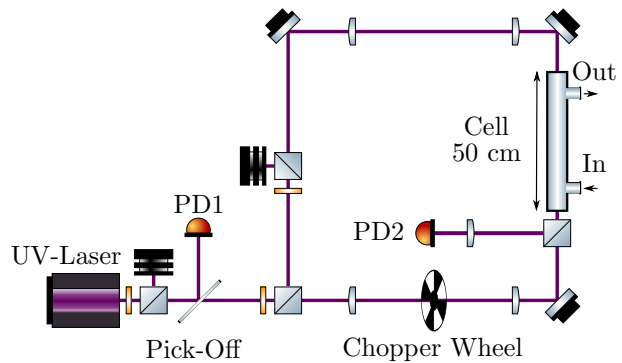

We employ Doppler-free saturated absorption spectroscopy \cite{Foot2005} as depicted in \FigRef{setup}. 
The laser source of the experiment is a frequency quadrupled titanium-sapphire laser tunable over the frequency range of the $\mathrm{A}\,^2\Sigma^+, (v\!=\!0) \leftarrow \mathrm{X}\,^2\Pi_{3/2}, (v\!=\!0)$ transition at \SI{226}{\nano\meter}.

The output power of the employed laser system is actively stabilized via a motorized halfwave-plate keeping the power on photodiode PD1 constant.
The laser beam is split into a pump and a probe beam. These two beams enter the \SI{50}{\centi\meter} long through flow spectroscopy cell in counter-propagating configuration with crossed linear polarizations. The probe beam is detected on photodiode PD2 (Thorlabs PDA25K2 - GaP Switchable Gain Amplified Detector).

The laser power was set to \SI{4}{\milli\watt} for the pump beam and \SI{0.4}{\milli\watt} for the probe beam. Both beams are expanded, the pump beam to approximately \SI{3.5}{\milli\meter} and the probe beam to \SI{2.8}{\milli\meter}, to ensure that transient time broadening can be neglected. The beam diameters of the collimated beams were measured direct in front of the cell entrance. The slightly larger pump beam ensures a homogeneous illumination for the probe beam.

The lamb-dip signal is not visible without employing a lock-in amplifier to increase the signal to noise ratio. Therefore, the amplitude of the pump beam is modulated at a frequency of \SI{9.8}{\kilo\hertz}. An  additional limitation is the frequency jitter of the UV-Laser occurring when the laser is scanned. This would lead to blurring and broadening of spectra which are averaged over several laser scans. To avoid this jitter the fundamental infrared beam of the quadrupled laser has to be stabilized. We employ an ultra-low expansion cavity to stabilize a master laser using the Pound-Drever-Hall technique \cite{Drever1983}. The  laser is send on a transfer cavity and its length is stabilized to the master laser with the same technique using a piezo. The fundamental beam of the quadrupled Ti:Sa can then be send on the transfer cavity. To stabilize the fundamental RF sidebands at two different frequencies are added using an electro-optic modulator (EOM). The sidebands needed to generate the Pound-Drever-Hall lock error signal are at a fixed frequency of \SI{15}{\mega\hertz}. To be able to lock the laser at any desired frequency the second sideband frequency is tunable from $\Delta\omega/2\pi=\SI{50}{\mega\hertz}$ to $\Delta\omega/2\pi=\SI{500}{\mega\hertz}$. The fixed sidebands and lock-sidebands are generated with different RF generators and added together by a power combiner. The lock-sidebands therefore also receive the  fixed \SI{15}{\mega\hertz} sidebands, so that it is possible to lock the laser to them.

Since nitric oxide dissociates rapidly under UV radiation \cite{ Minschwaner2001,Schellander2020} it is not possible to acquire comparable sets of data with a sealed spectroscopy cell. Therefore, the employed cell is connected to a vacuum pump system comprising a membrane and turbo pump to realize a controlled through flow system.
Between the outlet of the cell and the turbo pump, a choke system is installed limiting the suction power of the turbo pump by adjusting two butterfly valves. First the pressure is coarsely adjusted with the choke system. A constant pressure and flow is then ensured by a mass flow controller (MFC) in front of the cell. It is set so, to obtain a minimal pressure difference between the two gauges monitoring the pressure. They  are positioned directly at the inlet and outlet of the cell. Once the desired pressure is reached and the system is in equilibrium it is stable without the need of further adjustments on the MFCs or choke. The flow is regulated to an accuracy of \SI{0.2}{\percent} to \SI{1.0}{\percent} depending on the set value. The mean pressure was set to be around \SI{0.0230}{\milli\bar} for all datasets presented. 

The procedure to record a spectrum as shown in \FigRef{Traces} is as follows. First a continuous flow of NO is set and the laser is tuned to the coarse frequency of the desired rotational transition by referring to a wavemeter.
As a next step, the laser is scanned over a frequency range spanning several \si{\giga\hertz} to find the absorption profile of the corresponding spectral line. Since the experiment is conducted at room temperature, rotational lines are broadened by the Doppler-effect resulting in a linewidth of around \SI{3}{GHz}.
The laser scan width is then lowered stepwise to position the scan around the center of the absorption feature. For stronger rotational transitions the Lamb-dip is already visible without averaging the lock-in signal. This is beneficial for adjusting the beam overlap to achieve maximum signal strength while minimizing any reflection of the pump beam onto the detector PD2.
The laser can now be locked to the approximate position of the Lamb-dip or if it is not visible to the center of the Doppler-broadend absorption line by tuning the frequency of the lock-sidebands accordingly. In any case, before the actual measurement a coarse EOM-scan is performed.
Then the laser frequency is changed in steps of \SI{2}{\mega\hertz} up to a detuning of \SI{-80}{\mega\hertz}. For each frequency step a full integration cycle of the lock-in amplifier is collected. For coarse scans the time constant of the  lock-in amplifier is set to \SI{1}{\second}. If the lamb dip is visible on the coarse scan, a fine scan is performed. Otherwise the procedure is repeated for a different or larger frequency interval. For finer scans the frequency step size of the laser is set to \SI{0.4}{\mega\hertz} and the time constant of the lock-in amplifier to \SI{10}{\second}. 
\section{Results}\label{Sec3}

\begin{figure*}
    \centering
    \begin{tikzpicture}
    \node[anchor=south west] (image) at (0,0){\includegraphics[scale=1.0]{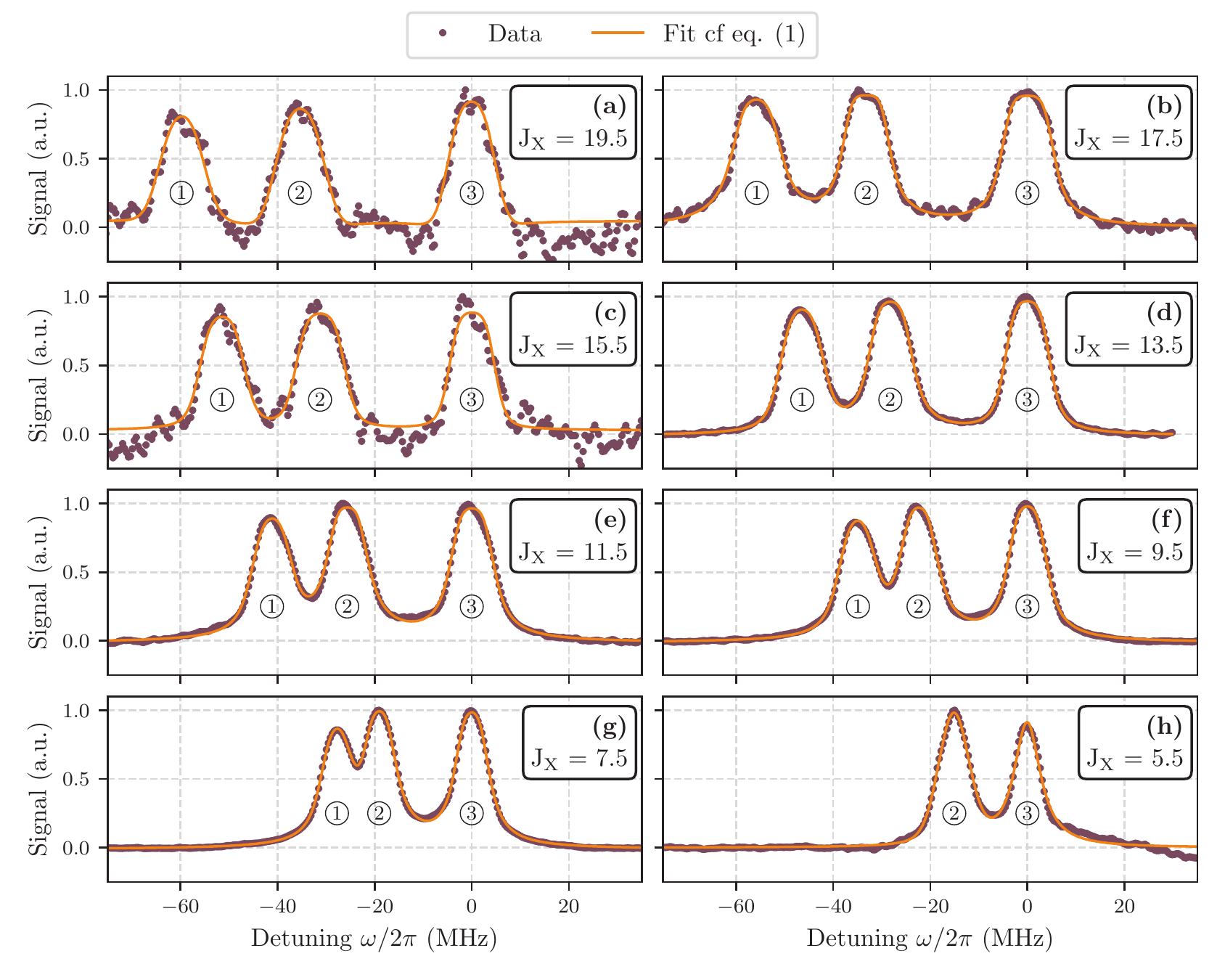}};
    \end{tikzpicture}
    \caption{\label{Traces} Selection of measured hyperfine spectra for different total angular momenta between (a) $J_\mathrm{X}=19.5$ and  (h) $5.5$. The data is shown as violet dots and was fitted with a Voigt profile of the form given in \EqnRef{VoigtFit}, shown as orange lines.}
\end{figure*}

\begin{figure}[htpb]
    \centering
    \begin{tikzpicture}
    \node[anchor=south west] (image) at (0,0){\includegraphics[scale=1.0]{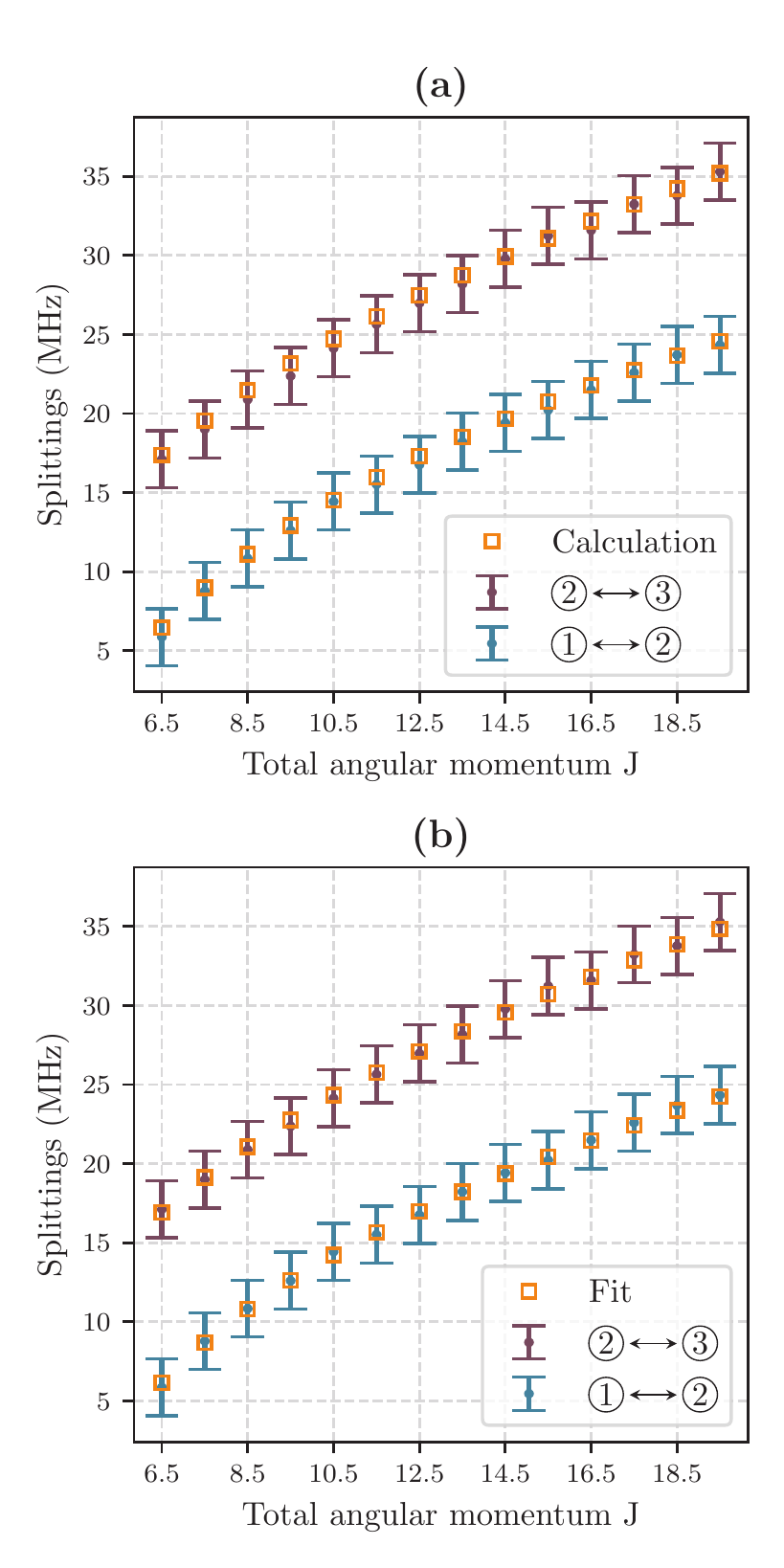}};
    \end{tikzpicture}
    \caption{\label{Splittings} (a) Comparison between measured and calculated hyperfine splittings. The measured splittings were calculated from the fit parameters of the Voigt-fit shown in \FigRef{Traces}. The errorbars include the estimated error of the laser lock and one standard deviation of the fit error. (b) Experimentally measured splittings as shown in (a) compared to data calculated from the newly fitted hyperfine constants for the $\mathrm{A} ^2\Sigma^+$ state. }
\end{figure}

The measurements on the $\mathrm{oP_{12ee}}$ branch yield Lamb-dip spectra for the total angular momenta $J_\mathrm{X}=1.5$ to $19.5$. The corresponding total angular momentum of the excited state is given by $J_\mathrm{A}=J_\mathrm{X}-1$. A selection of the recorded spectra is shown in \FigRef{Traces}. Peak \NiceCircleNumber{3} was set to zero detuning so that all other frequencies are given relative. In the range $J_\mathrm{X}=6.5$ to $19.5$ three individual hyperfine transitions (Lamb-dips) were resolved. With decreasing $J_\mathrm{X}$ the splitting between the transitions decreases. In \FigRef{Traces}{h} lines \NiceCircleNumber{1} and \NiceCircleNumber{2} have merged together so that only a single splitting can be determined. Therefore data for $J<6.5$ was not used for further evaluation.
At a later point, comparison of the data to a calculated spectrum will show that the decrease of the line splittings is the expected behavior, see \FigRef{Splittings}.

Since the beam overlap changed in between individual measurements and the mode of the UV-Laser is not  gaussian, the linewidth of the data will not be discussed. Our evaluation will focus on the splittings between the individually resolved lines instead. These are due to the energy structure of the molecule and not influenced by parameters like pressure, laser power or beam overlap. 
To determine the splitting between the measured hyperfine transitions the data was fitted with a Voigt-profile in a Lambert-Beer envelope after baseline and offset correction. The corresponding function is given in \EqnRef{VoigtFit} and depicted in \FigRef{Traces} as orange line,
\begin{equation}
        V(z) = A\cdot \left(1-\exp{\left(-1\cdot \sum_{k=1}^{3} \mathrm{Re}\left[ w_k(z(f))\right]\right)}\right).
        \label{VoigtFit}
\end{equation}
Here $A$ is the overall amplitude and $w_k(z)$ the Fadeeva function
\begin{equation}
    w_k(z)=\exp{\left(-z^2\right)}\left(1+\frac{2i}{\sqrt{\pi}}\int_0^z\exp{\left(t^2\right)}\mathrm{dt}\right).
    \label{Fadeeva}
\end{equation}
The argument $z$ of the normalized Fadeeva function is given by
\begin{equation}
    z(f) = \frac{f-f_0+i \gamma}{\sqrt{2}\sigma}
    \label{FadeevaArg}.
\end{equation}
It includes most of the remaining fit parameters, which are the frequency position of the maximum $f_0$, the Lorentzian linewidth $\gamma$ and the Doppler width $\sigma$ of the Voigt profile. The sum takes into account the number of peaks in the spectrum and $f$ is the measured frequency. From the fitparameter $f_0$ the splitting between the fitted lines can be calculated and compared to theoretically calculated values.

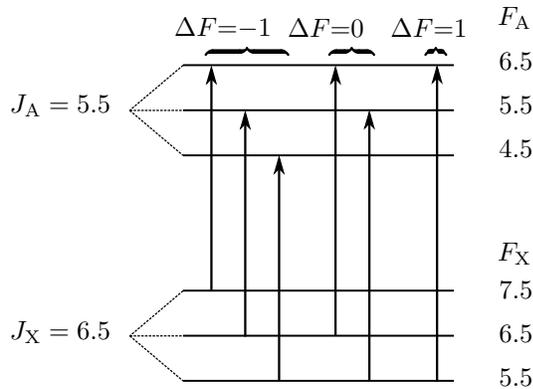
\begin{figure}[htbp]
 \centering
 \raisebox{-2em}{%
 \begin{tikzpicture}
        \node[anchor=south west] (image) at (0,0){\includegraphics[scale=0.6]{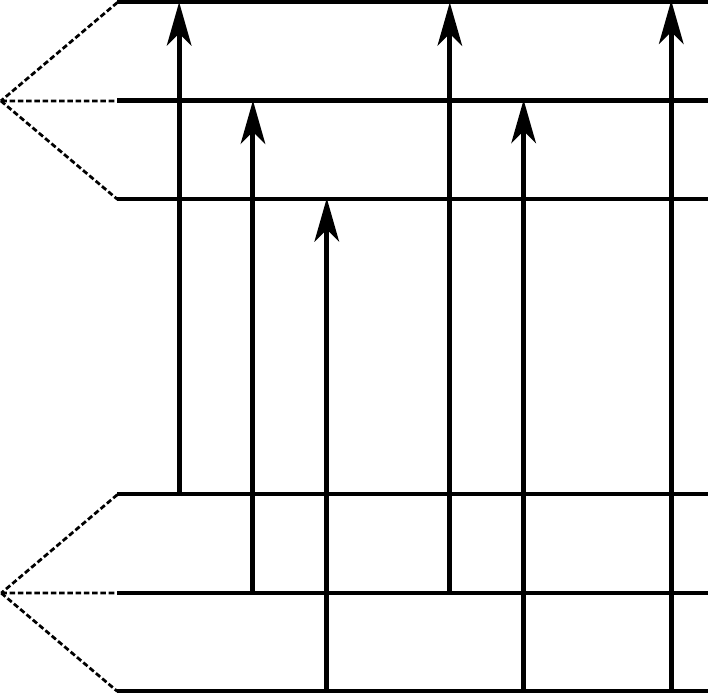}};
    %
    \begin{scope}[
    x={($0.1*(image.south east)$)},
    y={($0.1*(image.north west)$)}]
    %
    \node[text=black, anchor=east] at (-0,1.75) {$J_\mathrm{X}=6.5$};
    \node[text=black, anchor=east] at (-0,8.5) {$J_\mathrm{A}=5.5$}; 
    \node[text=black, anchor=south] at (11.5,10.5) {$F_\mathrm{A}$};
    \node[text=black, anchor=east] at (12.3,9.8) {$6.5$};
    \node[text=black, anchor=east] at (12.3,8.5) {$5.5$};
    \node[text=black, anchor=east] at (12.3,7.2) {$4.5$};
    \node[text=black, anchor=south] at (11.5,3.5) {$F_\mathrm{X}$};
    \node[text=black, anchor=east] at (12.3,3.0) {$7.5$};
    \node[text=black, anchor=east] at (12.3,1.7) {$6.5$};
    \node[text=black, anchor=east] at (12.3,0.4) {$5.5$};
    \draw [decorate, decoration = {calligraphic brace}, line width=1.75pt] (2.5,10) -- (4.9,10);
    \draw [decorate, decoration = {calligraphic brace}, line width=1.75pt] (6.1,10) -- (7.4,10);
    \draw [decorate, decoration = {calligraphic brace}, line width=1.75pt] (8.9,10) -- (9.5,10);
    \node[text=black, anchor=center] at (3.0,10.8) {$\thickmuskip=0mu \Delta F=-1$};
    \node[text=black, anchor=center] at (6.0,10.8) {$\thickmuskip=0mu \Delta F=0$};
    \node[text=black, anchor=center] at (9.0,10.8) {$\thickmuskip=0mu \Delta F=1$};
    \end{scope}
    \end{tikzpicture}
    }
    \caption{\label{DeltaF} Schematic of the of the six expected hyperfine transitions for an exemplary rotational transition of the $\mathrm{oP_{12ee}}$ branch. The energy splittings and arrows are not to scale.}
\end{figure}
As already mentioned the hyperfine structure is only partially resolved. \FIGRef{DeltaF} shows a schematic level scheme for all hyperfine transitions of a single rotational transition of the measured P-branch. According to the dipole selection rule for hyperfine transitions $\Delta F = \pm 1,0$, one expects a total of six hyperfine transitions to show up: Three transitions with $\Delta F=-1$, two transitions with  $\Delta F=0$ and one with  $\Delta F=+1$. To assign which of the six possible transitions were resolved, a full spectrum of the NO $\gamma$-band has been calculated using the software \emph{pgopher} \cite{Western2017}. The calculation is based on the constants given in \TblRef{Constants}.
\begin{table}[htbp]
    \caption{\label{Constants} List of constants used for the calculation with \emph{pgopher}. All values are given in MHz.}
        \begin{center}
            \begin{tabular}{c c c l}
                & $\mathrm{X} ^2\Pi$& \\ [0.5ex]
                \hline 
                Constant & Value &  Reference \\ [0.5ex] 
                \hline \hline
                $B$ & 50 848.130 72(18) & Varberg et al. \cite{Varberg1999} \\ 
                $D$ & 0.164 141 19(31) & .\\ 
                $H$ & 3.774(15) $\times 10^{-8}$ & .\\ 
                $A$ & 3 691 813.855(12) & Varberg et al. \cite{Varberg1999} \\ 
                $A_D$ & 5.372(38) & Danielak et al. \cite{Danielak1997}\\ 
                $\gamma$ & -193.987 9(77) & Varberg et al. \cite{Varberg1999} \\ 
                $\gamma_D$ & 0.001 582 2(70) & .\\ 
                $p$ & 350.405 443(91) & .\\ 
                $p_D$ & 3.78(18) $\times 10^{-5}$ & .\\ 
                $q$ & 2.822 100(51) & .\\ 
                $q_D$ & 4.370(38) $\times 10^{-5}$ & .\\ 
                $q_H$ & -8.6(25) $\times 10^{-10}$ & .\\ 
                $a$ & 84.203 78(76) & .\\ 
                $b_F$ & 22.379 2(28) & .\\ 
                $b$ & 42.006 5(38) & .\\ 
                $c$ & -58.882 0(32) & .\\ 
                $d$ & 112.597 18(13) & .\\ 
                $d_D$ & 1.10(23) $\times 10^{-4}$ &  .\\ 
                $C_I$ & 0.012347(52) & .\\ 
                $C^\prime_I$ & 0.007 37(36) &  .\\ 
                $eQq_0$ & -1.856 71(26) &  .\\ 
                $eQq_2$ & 23.114 7(83) & Varberg et al. \cite{Varberg1999} \\  [1.5ex]
                & $\mathrm{A} ^2\Sigma^+$& \\  [0.5ex] 
                \hline
                Constant & Value &  Reference \\ [0.5ex] 
                \hline \hline
                $T_0^A-T_0^X$ & 1 323 308 163(60) & Danielak et al. \cite{Danielak1997} \\
                $B$ & 59 545.297(293) & \cite{Brouard2012,Danielak1997}\\
                $D$ & 0.169572(236) & \cite{Brouard2012,Danielak1997}\\
                $\gamma$ & -80.34(16) &  \cite{Brouard2012,Danielak1997}\\
                $b_F$ & 43.52(30) & Brouard et al.  \cite{Brouard2012}  \\ 
                $b$ & 41.66(51) & .\\
                $c$ & 5.59(64) & .\\
                $eQq_0$ & -7.31(23) & Brouard et al.  \cite{Brouard2012}  \\ [1ex]
                \hline
                \end{tabular}
        \end{center}
\end{table}

It yields the six expected hyperfine transitions. However, there is a significant intensity difference between the transitions. The strongest three lines correspond to transitions with $\Delta F=-1$. They appear roughly 10 to 100 times stronger than the remaining two lines belonging to $\Delta F=0$ transitions and the line belonging to the $\Delta F=+1$ transition is significantly weaker than the  $\Delta F=0$ lines. This is the case because for $\Delta F=-1$ transitions only the orbital angular momentum of the electron has to be changed. For the $\Delta F=0,+1$ the nuclear spin has to change which leads to a smaller wavefunction overlap, thus a smaller transition dipole moment. Therefore we attribute the three measured transitions to $\Delta \mathrm{F}=-1$ transitions. The calculated and measured splittings are depicted in \SubFigRef{Splittings}{a}.
The violet and blue points represent measured data, the orange points correspond to the calculated splittings. The error bars include the fit error of the Voigt-fit and the estimated error of the laser stabilization which is the dominating contribution. It consists of the error of the master laser lock which is quadratically added to the error for the stabilization of the transfer cavity and the error of the lock of the laser itself. The total error due to the laser stabilization is around \SI{450}{\kilo\hertz} for the fundamental laser, thus it then is \SI{1.8}{\mega\hertz} in the UV. The lock errors were estimated via the root mean square of the noise of the error signal of the locked laser. In total the calculated and measured data show very good agreement.

In \SubFigRef{Splittings}{b} the measured data is compared to data calculated from the newly fitted constants. The fitting was performed employing a wrapper program for the command line version of \emph{pgopher} using a Levenberg-Marquardt algorithm for optimization. 
The hyperfine constants for the ground state of NO are very well determined \cite{Varberg1999} and were therefore kept fixed for the fitting procedure. The fine structure constants of the excited state were kept fixed as well. This leaves the three hyperfine constants $c$, $b$ and $eQq_0$ of the $A ^2\Sigma^+$ state to be fitted. From $b$ and $c$ the more convenient Fermi-contact constant $b_F$ can be calculated $b_F=b+\frac{c}{3}$. The corresponding values of the molecular constants determined by the fit are listed in table \ref{FitResults} and compared to the values given in \cite{Brouard2012}. In addition the column CPV gives the previously determined value which is closest to the result of this work. The agreement between this work and \cite{Brouard2012} is particularly good for the quadrupole constant $eQq_0$.
In contrast, the dipole-dipole interaction of the nuclear spin with the electron spin  $c$ deviates from the previous value but is within the combined error bounds. The nuclear spin electron spin interaction constant $b$ is just within the combined error bars.
The Fermi-contact constant is in less good agreement with \cite{Brouard2012} as the other constants and lies closest to the value determined by \cite{Gray1993}. Overall the constants determined in this work lie within the range of the previously determined values summarized in \cite{Brouard2012}.

\begin{table}
    \caption{\label{FitResults}Fit results in MHz for the fitted hyperfine constants $c, b, eQq_0$ compared to the values from \cite{Brouard2012} used for the initial calculation (see \SubFigRef{Splittings}{a} and to the closest previously determined value (CPV) with the corresponding source. An overview of the previously determined values for the respective constants is given in \cite{Brouard2012}.}
        \begin{center}
            \begin{tabular}{c c c c} \label{NewConst}
                Constant & This work &  Ref. \cite{Brouard2012} & CPV\\ [0.5ex] 
                \hline \hline
                $c$ & $5.29(53) $ & $5.59(64)$ & $5.59(64)$ \cite{Glendening1995}\\ 
                $b$ & $41.06(9) $ & $41.66(51)$ &  $41.66(51)$ \cite{Brouard2012}\\
                $eQq_0$ & $-7.31(12) $ & $-7.31(23)$ & $-7.31(23)$    \cite{Brouard2012}\\
                $b_F$ & $42.82(26) $ & $43.52(30)$ & $43.5(8)$             \cite{Gray1993}\\ [1ex] 
                \hline
                \end{tabular}
        \end{center}
\end{table}

\section{Summary}
Doppler-free saturated absorption spectroscopy has been employed to investigate hyperfine transitions between the $\mathrm{X} ^2\Pi_{3/2},v=0$ ground state manifold and the $\mathrm{A} ^2\Sigma^+,v=0$ state in nitric oxide. Spectra have been recorded for the $\mathrm{oP_{12ee}}$ branch with total angular momenta $J_\mathrm{X}=1.5-19.5$ of the ground state. The hyperfine structure resulting from the nuclear spin $I=1$ in nitrogen has been partially resolved. The extracted hyperfine splittings in the range of $J_\mathrm{X}=5.5-19.5$ were discussed and compared to calculations based on previously determined constants which show good agreement with the data.

Furthermore the hyperfine constants $b$, $c$ and $eQq_0$ for the $A ^2\Sigma^+$ state were fitted. Comparison of the newly determined constants show good agreement with the constants measured by Brouard et al. \cite{Brouard2012}.  
Overall the new values presented in this work fit into the range of previously published theoretical and experimental values.

For future experiments an even longer spectroscopy cell or a multipass-cell may be beneficial to measure at lower powers and pressures. This may increase the resolution further to resolve the weaker hyperfine lines. Measurement series for different laser powers and pressures may lead to a more precise determination of the yet only vaguely known saturation intensity and dynamical constants like excited state decay rates. 

Decreasing the pressure by two orders of magnitude allowed narrower spectroscopic lines but only for rotational energy levels close to the maximum of the rotational energy distribution in the ground state. For other lines the signal to noise ratio was not sufficient for data analysis. Therefore a full, consistent set of data could not be obtained at those parameters, limiting the measurements to pressure broadened lines.
\section*{Acknowledgements}
The authors thank the \Anonymize{European Commission (Grant Agreement No. 820393, MACQSIMAL)} for financial support. In addition the authors would like to thank \Anonymize{Professor Edward Grant and Professor Stephen Hogan} for fruitful discussions and valuable advice concerning the spectroscopic details of nitric oxide.
\section*{References}
\bibliographystyle{iopart-num}
\bibliography{literature}
\end{document}